\documentclass[prl,twocolumn]{revtex4-2}
\usepackage{amsmath}
\usepackage{amssymb}
\usepackage{graphicx}
\usepackage{braket}
\usepackage{stmaryrd}
\usepackage{hyperref}
\usepackage{color}
\usepackage{dsfont}

\hypersetup{
    colorlinks,
    citecolor=blue,
    linkcolor=blue,
    urlcolor=blue
}

\begin{document}
\title{Thermodynamic Correlation Inequality}
\author{Yoshihiko Hasegawa}
\email{hasegawa@biom.t.u-tokyo.ac.jp}
\affiliation{Department of Information and Communication Engineering, Graduate
School of Information Science and Technology, The University of Tokyo,
Tokyo 113-8656, Japan}
\date{\today}
\begin{abstract}
Trade-off relations
place fundamental limits on the operations that physical systems can perform. This Letter presents a
trade-off
relation that bounds the correlation function, which measures the relationship between a system's current and future states, in Markov processes. The obtained bound, referred to as the thermodynamic correlation inequality, states that the change in the correlation function has an upper bound comprising the dynamical activity, a
thermodynamic
measure of the activity of a Markov process. Moreover, by applying the obtained relation to the linear response function, it is demonstrated that the effect of perturbation can be bounded from above by the dynamical activity.

\end{abstract}
\maketitle

\textit{Introduction.---}
Trade-off relations
imply that there are impossibilities in the physical world that cannot be overcome by technological advances.
The most well-known example is the Heisenberg uncertainty relation \cite{Heisenberg:1927:UR,Robertson:1929:UncRel}, which establishes a limit on the precision of position-momentum measurement. 
The quantum speed limit is interpreted as an energy-time 
trade-off
relation and places a limit on the speed at which the quantum state can be changed \cite{Mandelstam:1945:QSL,Margolus:1998:QSL,Deffner:2010:GenClausius,Taddei:2013:QSL,DelCampo:2013:OpenQSL,Deffner:2013:DrivenQSL,Pires:2016:GQSL,OConnor:2021:ActionSL} (see \cite{Deffner:2017:QSLReview} for a review). 
It has many applications in quantum computation \cite{Lloyd:2000:CompLimit}, quantum communication \cite{Bekenstein:1981:InfoTransfer,Murphy:2010:QSLchain}, and quantum thermodynamics \cite{Deffner:2010:GenClausius}.
Recently, the concept of speed limit has also been considered in classical systems \cite{Shiraishi:2018:SpeedLimit,Ito:2018:InfoGeo,Ito:2020:TimeTURPRX}. 
In particular, the Wasserstein distance can be used to obtain the minimum entropy production required for a stochastic process to transform one probability distribution into another \cite{Dechant:2019:Wasserstein, Vu:2021:GeomBound,Nakazato:2021:Wasserstein,Dechant:2022:MarkovWasserstein,Vu:2022:OptimalTransportPRX}.
Moreover, the speed limit has been generalized to the time evolution of the observables \cite{Nicholson:2020:TIUncRel,GarciaPintos:2022:OSP,Hamazaki:2022:SL,Mohan:2022:OQSL,Hornedal:2022:KrylovSL}. 
A related principle, known as the thermodynamic uncertainty relation, was recently proposed in stochastic thermodynamics \cite{Barato:2015:UncRel,Gingrich:2016:TUP,Garrahan:2017:TUR,Dechant:2018:TUR,Terlizzi:2019:KUR,Hasegawa:2019:CRI,Hasegawa:2019:FTUR,Vu:2019:UTURPRE,Dechant:2020:FRIPNAS,Vo:2020:TURCSLPRE,Koyuk:2020:TUR,Pietzonka:2021:PendulumTURPRL,Erker:2017:QClockTUR,Brandner:2018:Transport,Carollo:2019:QuantumLDP,Liu:2019:QTUR,Guarnieri:2019:QTURPRR,Saryal:2019:TUR,Hasegawa:2020:QTURPRL,Hasegawa:2020:TUROQS,Sacchi:2021:BosonicTUR,Kalaee:2021:QTURPRE,Monnai:2022:QTUR}
(see \cite{Horowitz:2019:TURReview} for a review). This principle states that, for thermodynamic systems, higher accuracy can be achieved at the expense of higher thermodynamic costs. 
Recently, thermodynamic uncertainty relations have become a central topic in nonequilibrium thermodynamics; furthermore, their importance is also recognized from a practical standpoint because thermodynamic uncertainty relations can be used to infer entropy production without detailed knowledge of the system \cite{Li:2019:EPInference,Manikandan:2019:InferEPPRL,Vu:2020:EPInferPRE,Otsubo:2020:EPInferPRE}.

This Letter presents a 
trade-off
relation that confers bounds for the correlation function in Markov processes. 
The correlation function is a statistical measure that quantifies the correlation between the current state of a system and its future or past states. In a Markov process, the correlation function can be used to analyze the dependence of the current state on past states, and to identify any patterns in the system's behavior over time.
The correlation function provides spectral information through the Wiener-Khinchin theorem and plays a fundamental role in linear response theory \cite{Risken:1989:FPEBook}. 
Considering the significant role of the correlation function in stochastic processes, it is crucial to clearly illustrate its relationship with other physical quantities.
We derive the \textit{thermodynamic correlation inequality}, stating that the amount of correlation change has an upper bound that comprises the dynamical activity, which quantifies the activity of a system of interest. 
The derivation presented herein is based on considering the time evolution in a scaled path probability space \cite{Hasegawa:2023:BulkBoundaryBoundNC}, which can be regarded as a realization of bulk-boundary correspondence in Markov processes. 
By applying the H\"older inequality and a recently derived relation \cite{Hasegawa:2023:BulkBoundaryBoundNC}, the upper bound for the correlation function [Eq.~\eqref{eq:main_result_At}] is obtained.
The obtained bound exhibits unexpected generality; it holds for any Markov process
with an arbitrary time-independent transition rate
and can be generalized to multipoint correlation functions. 
The linear response function can be represented by the time derivative of the corresponding correlation function, as stated by the fluctuation--dissipation theorem. 
Upper bounds to the perturbation applied to the system are derived by applying
the thermodynamic correlation inequality
to the linear response function.

\begin{figure}
\includegraphics[width=8.5cm]{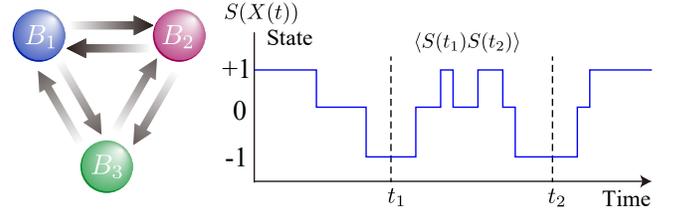} 
\caption{ 
Markov process with three states $\{B_1,B_2,B_3\}$. 
In this example,
we are interested in the correlation $\braket{S(t_1)S(t_2)}$,
where the score function $S(B_\nu)$
is specified by $S(B_1)=-1$, $S(B_2)=0$, and $S(B_3) = 1$. 
}
\label{fig:stochastic_process}
\end{figure}

\textit{Results.---}The thermodynamic correlation inequality is derived for a Markov process. 
Consider a Markov process with $N$ states, $\mathcal{B} \equiv \{B_1,B_2,\cdots,B_N\}$. 
Let $\{X(t)| t \ge 0\}$ be a collection of discrete random variables that take values in $\mathcal{B}$ (that is $X(t) \in \mathcal{B}$). 
Let $P(\nu;t)$ be the probability that $X(t)$ is $B_\nu$ at time $t$ and
$W_{\nu\mu}$ be the transition rate of $X(t)$ from $B_\mu$ to $B_\nu$.
The time evolution of $\mathbf{P}(t) \equiv [P(1;t),\ldots,P(N;t)]^\top$ is governed by the following master equation:
\begin{align}
\frac{d\mathbf{P}(t)}{dt}=\mathbf{W}\mathbf{P}(t),
\label{eq:master_equation_def}
\end{align}
where $\mathbf{W}\equiv \{W_{\nu\mu}\}$ and diagonal elements are defined as $W_{\nu\nu}\equiv -\sum_{\mu(\ne\nu)}W_{\mu\nu}$.
Next, we define a score function $S(B_\nu)$
that takes a state $B_\nu$ ($\nu \in \{1,2,\ldots,N\}$) and returns a real value of $(-\infty, \infty)$. 
Moreover, we define 
\begin{align}
    S_{\mathrm{max}}\equiv\max_{B\in\mathcal{B}}|S(B)|,
    \label{eq:Smax_def}
\end{align}
which is the maximum absolute value of the score function within $\mathcal{B}$. 
We also define another score function $T(B_\nu)$ similar to $S(B_\nu)$ and 
define $T_\mathrm{max}$ analogously. 
When it is clear from the context, we express $S(t) \equiv S(X(t))$ or $T(t) \equiv T(X(t))$ for simplicity. 
The correlation function 
$C(t) \equiv \braket{S(0)T(t)}$
is of interest, where 
\begin{align}
    \braket{S(0)T(t)}&=\sum_{\mu,\nu}T(B_{\nu})S(B_{\mu})P(\mu;0)P(\nu;t|\mu;0)\nonumber\\&=\mathds{1}\mathbf{T}e^{\mathbf{W}t}\mathbf{S}\mathbf{P}(0).
    \label{eq:correlation_def}
\end{align}
Here, $P(\nu;t|\mu;0)$ is the conditional probability that $X(t)=B_\nu$ given $X(0)=B_\mu$,
$\mathds{1}\equiv [1,1,\ldots,1]$ is the trace state, $\mathbf{S}\equiv \mathrm{diag}[S(B_1),\ldots,S(B_N)]$, and 
$\mathbf{T}\equiv\mathrm{diag}[T(B_{1}),\ldots,T(B_{N})]$.
The correlation function $C(t)$ has been extensively explored in the field of stochastic processes \cite{Masry:1972:UnitProcCov,Whitt:1991:SteadyStateSim}.
Recently, the correlation function was considered in the context of the quantum speed limit \cite{Mohan:2022:OQSL,Carabba:2022:CorQSLQJ}, which was obtained as a particular case of the speed limits on observables.
As an example of a classical system, a trichotomous process comprising three states $\mathcal{B}=\{B_1,B_2,B_3\}$ is shown in Fig.~\ref{fig:stochastic_process}.
$X(t)$ in this process exhibits random switching between $B_1$, $B_2$, and $B_3$.
For a trichotomous process, the score function is typically given by $S(B_1) = -1$, $S(B_2) = 0$, and $S(B_3) = 1$. 
To quantify the Markov process, 
we define the
time-integrated
dynamical activity $\mathcal{A}(t)$ as follows \cite{Maes:2020:FrenesyPR}:
\begin{align}
    \mathcal{A}(t)\equiv\int_{0}^{t}dt^{\prime}\sum_{\nu,\mu,\nu\neq\mu}P(\mu;t^{\prime})W_{\nu\mu}.
    \label{eq:dynamical_activity_def}
\end{align}
$\mathcal{A}(t)$ represents the average number of jumps during the interval $[0,t]$, and it quantifies the activity of the stochastic process. 
The dynamical activity plays a fundamental role in classical speed limits \cite{Shiraishi:2018:SpeedLimit} and thermodynamic uncertainty relations \cite{Garrahan:2017:TUR,Terlizzi:2019:KUR}.

In the Markov process, we obtain the upper bound of the correlation function $C(t)$. For $0\le t_1 < t_2$, we obtain the following bound:
\begin{align}
    \left|C(t_{1})-C(t_{2})\right|\le2S_{\mathrm{max}}T_{\mathrm{max}}\sin\left[\frac{1}{2}\int_{t_{1}}^{t_{2}}\frac{\sqrt{\mathcal{A}(t)}}{t}dt\right],
    \label{eq:main_result_At}
\end{align}
which holds for $0\le\frac{1}{2}\int_{t_{1}}^{t_{2}}\frac{\sqrt{\mathcal{A}(t)}}{t}dt\le\frac{\pi}{2}$. 
For $t_1$ and $t_2$ outside this range, the upper bound is 
$\left|C(t_{1})-C(t_{2})\right|\le2S_{\mathrm{max}}T_{\mathrm{max}}$,
which trivially holds true. 
Equation~\eqref{eq:main_result_At} is 
the \textit{thermodynamic correlation inequality} and
is the main result of this Letter. 
Note that all the quantities in Eq.~\eqref{eq:main_result_At} can be physically interpreted. 
A sketch of the proof of Eq.~\eqref{eq:main_result_At} is provided near the end of this Letter. 
Equation~\eqref{eq:main_result_At} holds for an arbitrary time-independent Markov process that starts from an arbitrary initial probability distribution with arbitrary score functions $S(B_\nu)$ and $T(B_\nu)$. 
Equation~\eqref{eq:main_result_At} states that higher dynamical activity allows the system to forget its current state quickly, which is in agreement with the intuitive notion.
In stochastic thermodynamics, entropy production plays a central role in thermodynamic inequalities. 
Entropy production measures the extent of irreversibility of a Markov process, whereas dynamical activity quantifies its intrinsic time scale.
Moreover, entropy production is not well defined for Markov processes that include irreversible transitions. By contrast, dynamical activity can be defined for any Markov process. This makes it particularly suitable for the correlation function, which needs to be calculated for any given Markov process.
A weaker bound can be obtained by using the thermodynamic uncertainty relation derived in a previous study~\cite{Hasegawa:2023:BulkBoundaryBoundNC} (see Ref.~\cite{Supp:2023:TCI} for details). 
\nocite{Wootters:1981:StatDist,Hamazaki:2023:VSL,Vo:2022:UKTURJPA,Nielsen:2011:QuantumInfoBook,LeCam:1973:Convergence,Sason:2016:DivIneqReview}
Let us consider particular cases of Eq.~\eqref{eq:main_result_At}. 
Taking $t_1=0$ and $t_2 = t$ with $t > 0$, Eq.~\eqref{eq:main_result_At} provides the upper bound for $|C(0)-C(t)|$:
\begin{align}
    \left|C(0)-C(t)\right|\le2S_{\mathrm{max}}T_{\mathrm{max}}\sin\left[\frac{1}{2}\int_{0}^{t}\frac{\sqrt{\mathcal{A}(t^{\prime})}}{t^{\prime}}dt^{\prime}\right],
    \label{eq:main_result_zero_tau}
\end{align}
where $0\le\frac{1}{2}\int_{0}^{t}\frac{\sqrt{\mathcal{A}(t^{\prime})}}{t^{\prime}}dt^{\prime}\le\frac{\pi}{2}$
(the saturating conditions are presented in Ref.~\cite{Supp:2023:TCI}).
Moreover, let $\epsilon$ be an infinitesimally small positive value. 
Substituting $t_1 = t - \epsilon$ and $t_2 = t$
into Eq.~\eqref{eq:main_result_At} and using the Taylor expansion for the sinusoidal function, we obtain
\begin{align}
    \left|\frac{dC(t)}{dt}\right|\le\frac{S_{\mathrm{max}}T_{\mathrm{max}}\sqrt{\mathcal{A}(t)}}{t}.
    \label{eq:main_result_derivative}
\end{align}
Equation~\eqref{eq:main_result_derivative} states that the absolute change of the correlation function is determined by the dynamical activity.
For $t\to 0$, the right side of Eq.~\eqref{eq:main_result_derivative} diverges to infinity. 
However, the derivative of $C(t)$ at $t=0$, represented as $\left|\partial_{t}C(t)\right|_{t=0}=\left|\mathds{1}\mathbf{T}\mathbf{W}\mathbf{S}\mathbf{P}(0)\right|$, is finite. This implies that the upper bound of Eq.~\eqref{eq:main_result_derivative} is not tight as $t$ approaches $0$.

As an intermediate step in the derivation of Eq.~\eqref{eq:main_result_zero_tau}, the following
inequality holds:
\begin{align}
   \left|C(0)-C(t)\right|\le2S_{\mathrm{max}}T_{\mathrm{max}}\sqrt{1-\eta(t)},
    \label{eq:main_result_eta}
\end{align}
where $\eta(t)$ is the Bhattacharyya coefficient between the path probabilities within $[0,t]$ having the transition rate matrix $\mathbf{W}$ and the null transition rate matrix $\mathbf{W}=0$. 
Since 
$\sqrt{1-\eta(t)}\le\sin\left[(1/2)\int_{0}^{t}\sqrt{\mathcal{A}\left(t^{\prime}\right)}/t^{\prime}\,dt^{\prime}\right]$,
Eq.~\eqref{eq:main_result_eta} is tighter than Eq.~\eqref{eq:main_result_zero_tau}.
The inequality of Eq.~\eqref{eq:main_result_eta} holds for any value of $t$ because the Bhattacharyya coefficient is always bounded between $0$ and $1$.
$\eta(t)$ can be computed as 
$\eta(t)\equiv\left(\sum_{\mu}P(\mu;0)\sqrt{e^{-t\sum_{\nu(\ne\mu)}W_{\nu\mu}}}\right)^{2}$, 
which can be represented by quantities of the Markov process \cite{Supp:2023:TCI}. 
Note that $\eta(t)$ constitutes a lower bound in thermodynamic uncertainty relations \cite{Hasegawa:2021:QTURLEPRL}. 
The term within the square root in $\eta(t)$ represents the survival probability that there is no jump starting from $B_\mu$. 
Therefore, when the activity of the dynamics is lower, the survival probability increases and, in turn, $\eta(t)$ yields a higher value. 
Although $\eta(t)$ has fewer physical interpretations than dynamical activity $\mathcal{A}(t)$, 
it has an advantage over Eq.~\eqref{eq:main_result_zero_tau} that 
the bound of Eq.~\eqref{eq:main_result_eta} holds for any value of $t$. 

Now, we comment on possible improvements and generalizations to the thermodynamic correlation inequality. 
The inequality can be tightened by replacing 
$S_\mathrm{max}T_\mathrm{max}$, included in Eq.~\eqref{eq:main_result_At}, with 
$\frac{1}{2}\left[\max_{B_{1},B_{2}\in\mathcal{B}}S(B_{1})T(B_{2})-\min_{B_{1},B_{2}\in\mathcal{B}}S(B_{1})T(B_{2})\right]$. 
In addition, it is also possible to consider the $J$-point correlation function ($J\ge 2$ is an integer), which serves as a generalization of the two-point correlation function discussed above. 
The results are presented in detail in Ref.~\cite{Supp:2023:TCI}.

\begin{figure}
\includegraphics[width=8.5cm]{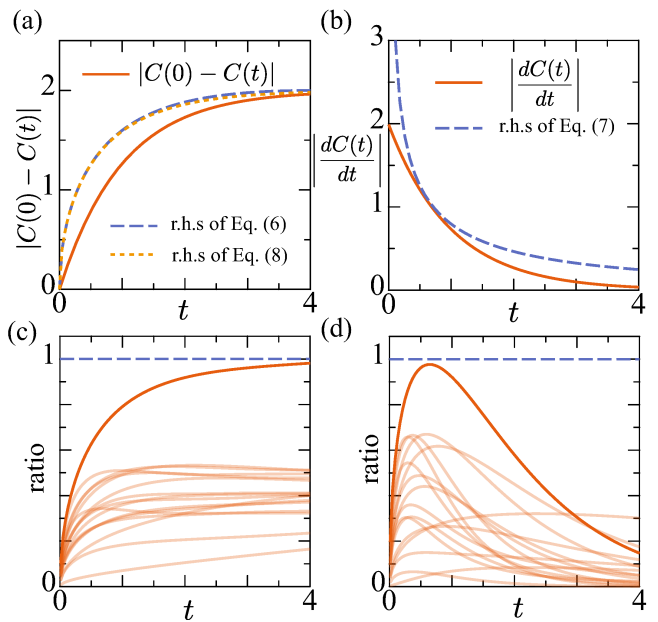} 
\caption{
(a) $|C(0)-C(t)|$ as a function $t$ for the two-state Markov process,
which is shown by the solid line. 
Its two upper bounds, the right-hand sides of Eqs.~\eqref{eq:main_result_zero_tau} and \eqref{eq:main_result_eta}, are depicted by dashed and dotted lines, respectively. 
(b) $|\partial_{t}C(t)|$ as a function $t$ for the two-state Markov process and its upper bound, which are shown by the solid and dashed lines, respectively. 
In the two-state Markov process in (a) and (b), the transition rate is $W_{12}=1$ and $W_{22}=-1$ ($0$ for the other entries), the initial distribution is $\mathbf{P}(0)=[0,1]^\top$, and the score function is $S(B_1)=-1$, $S(B_2)=1$,
$T(B_1)=-1$, and $T(B_2)=1$.
(c) Ratio 
$|C(0)-C(t)|/\left(2S_{\mathrm{max}}T_{\mathrm{max}}\sin\left[(1/2)\int_{0}^{t}\sqrt{\mathcal{A}\left(t^{\prime}\right)}/t^{\prime}\,dt^{\prime}\right]\right)$ 
as a function of $t$. 
The light solid lines are random realizations, whereas the dark solid line corresponds to the setting of (a). 
(d) Ratio 
$\left|\partial_{t}C(t)\right|/\left(S_{\mathrm{max}}T_{\mathrm{max}}\sqrt{\mathcal{A}(t)}/t\right)$
as a function of $t$. 
The light solid lines are random realizations and the dark solid line corresponds to the setting of (b). 
In (c) and (d), the ratios must not exceed $1$,
which is depicted by the dashed lines. 
For the random realizations in (c) and (d), we randomly determine the transition rate $\mathbf{W}$, the initial probability $\mathbf{P}(0)$,
and the score function $S(B_\nu)$ for $N=2,3,4$, where we use $T(B_\nu)=S(B_\nu)$ for $T(B_\nu)$. 
}
\label{fig:bound_plot}
\end{figure}

We perform numerical simulations to validate Eqs.~\eqref{eq:main_result_zero_tau}--\eqref{eq:main_result_eta}.
We prepare a two-state Markov process
($\mathcal{B} = \{B_1,B_2\}$)
and plot $|C(0)-C(t)|$ and $|\partial_{t}C(t)|$ as functions of $t$ in Figs.~\ref{fig:bound_plot}(a) and (b), respectively, by the solid lines (see the caption of Fig.~\ref{fig:bound_plot} for details).
In Fig.~\ref{fig:bound_plot}(a), 
we plot the right-hand sides of Eqs.~\eqref{eq:main_result_zero_tau}
and \eqref{eq:main_result_eta}, which are upper bounds of $|C(0)-C(t)|$, by the dashed and dotted lines, respectively. 
Furthermore, we plot the right-hand side of Eq.~\eqref{eq:main_result_derivative}, the upper bound of $|\partial_{t}C(t)|$, by the dashed line in Fig.~\ref{fig:bound_plot}(b).
From Fig.~\ref{fig:bound_plot}(a), we observe that Eq.~\eqref{eq:main_result_zero_tau} provides an accurate estimate of the upper bound. 
In this case, the difference between the two upper bounds given by Eqs.~\eqref{eq:main_result_zero_tau} and \eqref{eq:main_result_eta} is negligible. 
The upper bound shown in Fig.~\ref{fig:bound_plot}(b) becomes less tight for a large $t$, because the decay of the upper bound is approximately $O(t^{-1/2})$ whereas
the correlation function decays exponentially
in this model.
Next, we randomly generate Markov processes and verify whether the bounds hold for the random realizations (see the caption of Fig.~\ref{fig:bound_plot} for details). 
We calculate the ratio, the left-hand sides divided by the right-hand sides of Eqs.~\eqref{eq:main_result_zero_tau} and \eqref{eq:main_result_derivative}, in Figs.~\ref{fig:bound_plot}(c) and (d), respectively, by the light solid lines. 
The ratio should not exceed $1$, as indicated by the dashed lines. 
In Figs.~\ref{fig:bound_plot}(c) and (d), the dark solid lines correspond to the results shown in Figs.~\ref{fig:bound_plot}(a) and (b), respectively. 
All realizations are below $1$, which numerically verifies the bounds.

\textit{Linear response.---}The correlation function $C(t)$ is closely related to linear response theory \cite{Risken:1989:FPEBook}.
The correlation bounds in Eqs.~\eqref{eq:main_result_zero_tau} and \eqref{eq:main_result_derivative} are applied to the linear response theory \cite{Supp:2023:TCI}.
Suppose that a Markov process is in the steady state $\mathbf{P}_{st} = [P_{st}(1),\ldots,P_{st}(N)]^\top$, which satisfies $\mathbf{W}\mathbf{P}_{st} = 0$. 
A weak perturbation $\chi \mathbf{F} f(t)$ is applied to the master equation in Eq.~\eqref{eq:master_equation_def}, which is $\mathbf{W}\to \mathbf{W} + \chi \mathbf{F} f(t)$ in Eq.~\eqref{eq:master_equation_def}. Here,
$\chi$ denotes the perturbation strength satisfying
$0 < |\chi| \ll 1$, $\mathbf{F}$ is an $N\times N$ matrix, and $f(t)$ is arbitrary real function of time $t$. 
The probability distribution is expanded as $\mathbf{P}(t) = \mathbf{P}_{st} + \chi \mathbf{P}_1(t)$, where $\mathbf{P}_1(t)$ is the first-order correction to the probability distribution. 
By collecting the first-order contribution $O(\chi)$ in Eq.~\eqref{eq:master_equation_def}, $\mathbf{P}_1(t)$ is given by \cite{Supp:2023:TCI}
\begin{align}
    \mathbf{P}_{1}(t)=\int_{-\infty}^{t}e^{\mathbf{W}(t-t^{\prime})}\mathbf{F}\mathbf{P}_{st}f(t^{\prime})dt^{\prime}.
    \label{eq:P1_sol_main}
\end{align}
Let a score function $G(B_\nu)$ be considered. Define the expectation of $G(B_\nu)$ as 
$\braket{G}=\sum_{\nu}G(B_{\nu})P(\nu;t)=\mathds{1}\mathbf{G}\mathbf{P}(t)$,
where $\mathbf{G} \equiv \mathrm{diag}[G(B_1),\ldots,G(B_N)]$.
The change in $\braket{G}$ due to the perturbation, represented by $\Delta G\equiv\mathds{1}\mathbf{G}\mathbf{P}(t)-\mathds{1}\mathbf{G}\mathbf{P}_{st}$, is
$\Delta G(t)=\chi\int_{-\infty}^{\infty}R_{G}(t-t^{\prime})f(t^{\prime})dt^{\prime}$,
where $R_{G}(t)$ denotes the linear response function:
\begin{align}
    R_{G}(t)=\begin{cases}
\mathds{1}\mathbf{G}e^{\mathbf{W}t}\mathbf{F}\mathbf{P}_{st} & t\ge0\\
0 & t<0
\end{cases}.
\label{eq:linear_response_function_def}
\end{align}
In the linear response regime, any input-output relation can be expressed through $R_G(t)$. 
From Eq.~\eqref{eq:correlation_def}, the time derivative of $C(t)$ is
$\partial_{t}C(t)=\mathds{1}\mathbf{T}e^{\mathbf{W}t}\mathbf{W}\mathbf{S}\mathbf{P}_{st}$.
Comparing Eq.~\eqref{eq:linear_response_function_def} and $\partial_t C(t)$, when 
$\mathbf{G}=\mathbf{T}$
and $\mathbf{F} = \mathbf{W}\mathbf{S}$,
then $\partial_t C(t)$ provides the linear response function of Eq.~\eqref{eq:linear_response_function_def}, which is the statement of the fluctuation-dissipation theorem.

As a particular case, let us consider the pulse perturbation $f(t)=\delta(t)$, where $\delta(t)$ is the Dirac
delta function. 
This perturbation corresponds to the application of a sharp pulsatile perturbation at $t=0$.
Then the change in the expectation of 
$T(B_\nu)$
under the perturbation $\mathbf{F} = \mathbf{W}\mathbf{S}$, represented by
$\Delta T^{(p)}$, is  $\Delta T^{(p)}(t)=\chi\partial_{t}C(t)$
(the superscript (p) represents that it is the pulse response).
The correlation bound in Eq.~\eqref{eq:main_result_derivative} yields
\begin{align}
\left|\Delta T^{(p)}(t)\right|\le\chi S_{\mathrm{max}}T_{\mathrm{max}}\sqrt{\frac{\mathfrak{a}}{t}}\,\,\,\,(t>0),
    \label{eq:Delta_S_p_bound}
\end{align}
where $\mathfrak{a}$ is the dynamical activity $\mathfrak{a}\equiv\sum_{\nu,\mu,\nu\neq\mu}P_{st}(\mu)W_{\nu\mu}$ 
(note that $\mathcal{A}(t) = \mathfrak{a}t$ for the steady state).
Equation~\eqref{eq:Delta_S_p_bound} relates the dynamical activity to the effect of the pulse perturbation in the Markov process. 
The step response can be calculated similarly. 
We apply a constant perturbation switched on at $t=0$, which can be modeled by $f(t) = \Theta(t)$ with $\Theta(t)$ being the Heaviside step function. 
We obtain
$\Delta T^{(s)}(t)=\chi\int_{0}^{t}R_{T}(t^{\prime})dt^{\prime}=\chi\left(C(t)-C(0)\right)$,
which along with Eq.~\eqref{eq:main_result_zero_tau} yields the following bound:
\begin{align}
    |\Delta S^{(s)}(t)|\le2\chi S_{\mathrm{max}}T_{\mathrm{max}}\sin\left[\sqrt{\mathfrak{a}t}\right]\,\,\,\,(t>0).
    \label{eq:Delta_S_s_bound}
\end{align}
Equation~\eqref{eq:Delta_S_s_bound} holds for $0\le\sqrt{\mathfrak{a}t}\le\pi/2$. 
For $t$ outside this range, the trivial inequality 
$|\Delta S^{(s)}(t)|\le2\chi S_{\mathrm{max}}T_{\mathrm{max}}$
holds true.

\begin{figure}
\includegraphics[width=8.5cm]{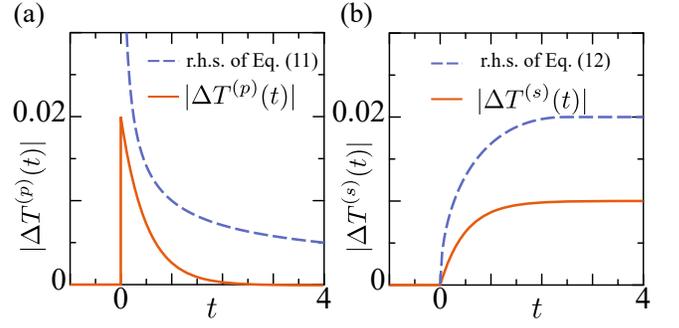} 
\caption{
(a) Response under the pulse perturbation,
where the solid line denotes 
$\Delta T^{(p)}(t)$
as a function of $t$ and the dashed line shows its upper bound [the right-hand side of Eq.~\eqref{eq:Delta_S_p_bound}]. 
(b) Response under the step perturbation,
where the solid line denotes 
$\Delta T^{(s)}(t)$
as a function of $t$ and the dashed line shows its upper bound [the right-hand side of Eq.~\eqref{eq:Delta_S_s_bound}]. 
In (a) and (b), the transition matrix is $W_{12}=W_{21}=1$ (the diagonal elements are $W_{11}=W_{22}=-1$), and 
the score functions are $S(B_1)=-1$, $S(B_2)=1$, $T(B_1)=-1$, and $T(B_2)=1$.
Additionally, 
the perturbation strength
$\chi$ is set to $0.01$.
}
\label{fig:linear_response_plot}
\end{figure}

We perform numerical simulations to validate Eqs.~\eqref{eq:Delta_S_p_bound} and \eqref{eq:Delta_S_s_bound}.
We prepare a two-state Markov process
($\mathcal{B} = \{B_1,B_2\}$)
and plot 
$|\Delta T^{(p)}(t)|$ and $|\Delta T^{(s)}(t)|$
as functions of $t$ in Figs.~\ref{fig:linear_response_plot}(a) and (b), respectively, by the solid lines (see the caption of Fig.~\ref{fig:linear_response_plot} for details).
We plot their upper bounds by the dashed lines.
From Figs.~\ref{fig:linear_response_plot}(a) and (b),
we can observe that the bounds are satisfied for both systems. 
As $t$ increases, the upper bound loosens in Fig.~\ref{fig:linear_response_plot}(a). This is because the upper bound decays at $O(t^{-1/2})$, whereas the decay rate of 
$|\Delta T^{(p)}(t)|$
is exponential.
In Fig.~\ref{fig:linear_response_plot}(b),
at $t=4$, there is a two-fold gap between the bound and 
$\Delta T^{(s)}(t)$;
however, if the bound is halved, the bound is invalid.

\textit{Conclusion.---}This letter presents the relation between the correlation function and dynamical activity in the Markov process. 
The obtained bounds hold for an arbitrary time-independent transition rate starting from an arbitrary initial distribution. 
By applying the obtained bounds to the linear response theory, we demonstrated that the effect of perturbations on a steady-state system is bounded by the dynamical activity. The findings herein can potentially enhance our understanding of nonequilibrium dynamics, as the correlation function plays a fundamental role in thermodynamics.

\textit{Appendix: Derivation.---}Here, we provide a sketch of proof of Eqs.~\eqref{eq:main_result_At} and \eqref{eq:main_result_eta}. For details of the derivation, refer to Ref.~\cite{Supp:2023:TCI}. 

Let $p(x;t)$ be the general probability distribution of $x$ at time $t$ ($x$ is an arbitrary random variable). 
Let $F(x)$ be an observable of $x$ and $\braket{F}_t\equiv \sum_x F(x)p(x;t)$ be the expectation of $F(x)$ at time $t$. 
From the H\"older inequality, the following relation holds:
\begin{align}
    \left|\braket{F}_{t_{1}}-\braket{F}_{t_{2}}\right|\le2F_{\max}\mathrm{TVD}\left(p(x;t_{1}),p(x;t_{2})\right),
    \label{eq:F_Holder_ineq}
\end{align}
where $F_{\max} \equiv \max_x |F(x)|$ and $\mathrm{TVD}(\cdot,\cdot)$ is the total variation distance:
\begin{align}
    \mathrm{TVD}(p(x;t_{1}),p(x;t_{2}))\equiv\frac{1}{2}\sum_{x}|p(x;t_{1})-p(x;t_{2})|.
    \label{eq:TVD_def}
\end{align}
The speed limit relations are conventionally concerned with the time evolution of $\mathbf{P}(t)$. 
In contrast, 
we consider the time evolution of the path probability in Eq.~\eqref{eq:F_Holder_ineq},
which was previously studied~\cite{Hasegawa:2023:BulkBoundaryBoundNC}.
The final time $\tau > 0$ of the Markov process is first fixed. 
Let $\omega_{t}\equiv[X(t^{\prime})]_{t^{\prime}=0}^{t^{\prime}=t}$ be the trajectory of a Markov process within the time interval $[0,t]$ ($0\le t \le \tau$) and let $\mathcal{P}(\omega_t;\mathbf{W})$ be the path probability (path integral) with the transition rate $\mathbf{W}$. 
We can not substitute $\mathcal{P}(\omega_t;\mathbf{W})$ into Eq.~\eqref{eq:F_Holder_ineq} because the 
size
of $\omega_t$ is different for different $t$. 
Therefore, we introduce the scaled process \cite{Hasegawa:2023:BulkBoundaryBoundNC}:
\begin{align}
    \mathcal{Q}(\omega_{\tau};t)\equiv\mathcal{P}\left(\omega_{\tau};\frac{t}{\tau}\mathbf{W}\right).
    \label{eq:Q_def}
\end{align}
In Eq.~\eqref{eq:Q_def}, $\mathcal{Q}(\omega_{\tau};t)$ is the path probability of ``scaled'' process;
the scaled process is the same as the original process, except for its time scale. 
In the scaled process, the transition rate is $(t/\tau)\mathbf{W}$, which is $t/\tau$ times faster than the original process. 
Therefore, the information at time $t$ in the original process with the transition rate $\mathbf{W}$ can be obtained at time $\tau$ in the scaled process with the transition rate $(t/\tau)\mathbf{W}$.
The total variation distance admits the following upper bound:
\begin{align}
    &\mathrm{TVD}(\mathcal{Q}(\omega_{\tau};t_{1}),\mathcal{Q}(\omega_{\tau};t_{2}))\nonumber\\
    &\leq\sqrt{1-\mathrm{Bhat}(\mathcal{Q}(\omega_{\tau};t_{1}),\mathcal{Q}(\omega_{\tau};t_{2}))^{2}}.
    \label{eq:TVD_Bhat_ineq}
\end{align}
Using the results of Ref.~\cite{Hasegawa:2023:BulkBoundaryBoundNC},
the following relation holds for $0\le t_1 < t_2 \le \tau$:
\begin{align}
    \frac{1}{2}\int_{t_{1}}^{t_{2}}\frac{\sqrt{\mathcal{A}(t)}}{t}dt\ge\arccos\mathrm{Bhat}\left(\mathcal{Q}\left(\omega_{\tau};t_{1}\right),\mathcal{Q}\left(\omega_{\tau};t_{2}\right)\right).
    \label{eq:geodesic_Bhat}
\end{align}
Substituting Eq.~\eqref{eq:geodesic_Bhat} into Eq.~\eqref{eq:TVD_Bhat_ineq}, we obtain
\begin{align}
    \mathrm{TVD}\left(\mathcal{Q}(\omega_{\tau};t_{1}),\mathcal{Q}(\omega_{\tau};t_{2})\right)\le\sin\left[\frac{1}{2}\int_{t_{1}}^{t_{2}}\frac{\sqrt{\mathcal{A}(t)}}{t}dt\right].
    \label{eq:TVD_upperbound}
\end{align}

Consider an observable $\mathcal{C}(\omega_\tau)$, defined as
\begin{align}
    \mathcal{C}(\omega_\tau) \equiv S(X(0))T(X(\tau)).
    \label{eq:C_obs_def}
\end{align}
Then the expectation of $\mathcal{C}(\omega_\tau)$ with respect to $\mathcal{Q}(\omega_\tau;t)$ yields the correlation, i.e., 
$\braket{\mathcal{C}(\omega_{\tau})}_{t}\equiv\sum_{\omega_{\tau}}\mathcal{Q}(\omega_{\tau};t)\mathcal{C}(\omega_{\tau})=\braket{S(X(0))T(X(t))}$.
Combining Eqs.~\eqref{eq:F_Holder_ineq} and \eqref{eq:TVD_upperbound}, and
considering $\mathcal{C}(\omega_\tau)$ for the observable in Eq.~\eqref{eq:F_Holder_ineq}, we obtain
\begin{align}
    \left|\braket{\mathcal{C}}_{t_{1}}-\braket{\mathcal{C}}_{t_{2}}\right|\le2\mathcal{C}_{\max}\sin\left[\frac{1}{2}\int_{t_{1}}^{t_{2}}\frac{\sqrt{\mathcal{A}(t)}}{t}dt\right],
    \label{eq:main_result_1_pre}
\end{align}
which leads to the main result of Eq.~\eqref{eq:main_result_At}.

Let us derive the bound of Eq.~\eqref{eq:main_result_eta},
which can be obtained as an intermediate step in the above derivation. 
Instead of using Eq.~\eqref{eq:TVD_upperbound} for deriving the bound, 
we employ Eq.~\eqref{eq:TVD_Bhat_ineq} with $t_1 = 0$ and $t_2 =\tau$. 
The Bhattacharyya coefficient yields $\mathrm{Bhat}\left(\mathcal{Q}\left(\omega_{\tau};0\right),\mathcal{Q}\left(\omega_{\tau};\tau\right)\right)=\sum_{\mu}P(\mu;0)\sqrt{e^{-\tau\sum_{\nu(\ne\mu)}W_{\nu\mu}}}$ (see Ref.~\cite{Supp:2023:TCI} for details), 
which provides the bound in Eq.~\eqref{eq:main_result_eta}.

\begin{acknowledgments}
This work was supported by JSPS KAKENHI Grant Number JP22H03659.
\end{acknowledgments}

\end{document}

\title{Supplementary Material for ``Thermodynamic Correlation Inequality''}
\author{Yoshihiko Hasegawa}
\email{hasegawa@biom.t.u-tokyo.ac.jp}
\affiliation{Department of Information and Communication Engineering, Graduate
School of Information Science and Technology, The University of Tokyo,
Tokyo 113-8656, Japan}

\maketitle
This supplementary material describes the calculations introduced in the main text. The equations and figure numbers are prefixed with ``S'' [for example, Eq.~(S1) or Fig.~S1]. Numbers without this prefix [for example, Eq.~(1) or Fig.~1] refer to items in the main text.

\section{Derivation of Eqs.~\mainUresultUAt{} and \mainUresultUeta{}
\label{sec:Appendix_main_At_proof}}

Here, we derive Eq.~\mainUresultUAt{}. 
Let $\mathcal{F}$ be an arbitrary Hermitian operator and $\braket{\mathcal{F}}_{t}\equiv\mathrm{Tr}[\rho(t)\mathcal{F}]$,
where $\rho(t)$ is a density operator.
The following relation has recently been used in the field of quantum speed limits \cite{Hamazaki:2022:SL,Mohan:2022:OQSL}:
\begin{align}
    \left|\braket{\mathcal{F}}_{t_{1}}-\braket{\mathcal{F}}_{t_{2}}\right|&=\left|\mathrm{Tr}\left[\mathcal{F}(\rho(t_{1})-\rho(t_{2}))\right]\right|\nonumber\\&\le\left\Vert \mathcal{F}\right\Vert _{\mathrm{op}}\left\Vert \rho(t_{1})-\rho(t_{2})\right\Vert _{\mathrm{tr}}\nonumber\\&=2\left\Vert \mathcal{F}\right\Vert _{\mathrm{op}}\mathrm{TD}\left(\rho(t_{1}),\rho(t_{2})\right),
    \label{eq:Operator_ineq}
\end{align}
where $\left\Vert \cdot\right\Vert _{\mathrm{op}}$ and $\left\Vert \cdot\right\Vert _{\mathrm{tr}}$ denote the operator norm [Eq.~\eqref{eq:Shatten_inf}] and trace norm [Eq.~\eqref{eq:Shattan_one}], respectively, and $\mathrm{TD}(\cdot,\cdot)$ is the trace distance [Eq.~\eqref{eq:trace_dist_def}].
The transformation 
from the first line to the second line
of Eq.~\eqref{eq:Operator_ineq} comes from the H\"older inequality 
 (see Eq.~\eqref{eq:Holder_inf_1}).

Considering Eq.~\eqref{eq:Operator_ineq} for the classical probability space, we assume that the two density operators have only diagonal elements:
\begin{align}
    \rho(t) = \sum_x p(x; t) \ket{x}\bra{x},
    \label{eq:rho_diagonal_def}
\end{align}
where $p(x;t)$ is a probability distribution and $\{\ket{x}\}_x$ constitutes an orthonormal basis. 
By calculating the trace distance for Eq.~\eqref{eq:rho_diagonal_def}, $\mathrm{TD}(\rho,\sigma)$ is reduced to the total variation distance [Eq.~\eqref{eq:TVD_def}]:
\begin{align}
    \mathrm{TD}(\rho(t_{1}),\rho(t_{2}))=\mathrm{TVD}(p(x;t_{1}),p(x;t_{2})).
    \label{eq:TD_TVD_rel}
\end{align}
Moreover, assuming that the observable $\mathcal{F}$ in Eq.~\eqref{eq:Operator_ineq} is diagonal with respect to $\ket{x}$:
\begin{align}
    \mathcal{F}=\sum_{x}F(x)\ket{x}\bra{x},
    \label{eq:F_diagonal}
\end{align}
where $F(x)$ is a real function. 
Then the left-hand side of Eq.~\eqref{eq:Operator_ineq} becomes
\begin{align}
    \left|\braket{\mathcal{F}}_{t_{1}}-\braket{\mathcal{F}}_{t_{2}}\right|=\left|\braket{F}_{t_{1}}-\braket{F}_{t_{2}}\right|=\left|\sum_{x}F(x)p(x;t_{1})-\sum_{x}F(x)p(x;t_{2})\right|,
    \label{eq:LHS_classical}
\end{align}
where $\braket{F}_t = \sum_x F(x)p(x;t)$, as defined in the main text. 
Subsequently, from Eqs.~\eqref{eq:TD_TVD_rel} and \eqref{eq:LHS_classical},
Eq.~\eqref{eq:Operator_ineq} reduces to
\begin{align}
    \left|\braket{F}_{t_{1}}-\braket{F}_{t_{2}}\right|\le2F_{\max}\mathrm{TVD}\left(p(x;t_{1}),p(x;t_{2})\right),
    \label{eq:F_Holder_ineq}
\end{align}
where $\left\Vert \mathcal{F}\right\Vert _{\mathrm{op}} = F_{\max} = \max_x |F(x)|$. 
Equation~\eqref{eq:F_Holder_ineq} corresponds to Eq.~\FUHolderUineq{} in the main text.

\begin{figure*}
\includegraphics[width=16cm]{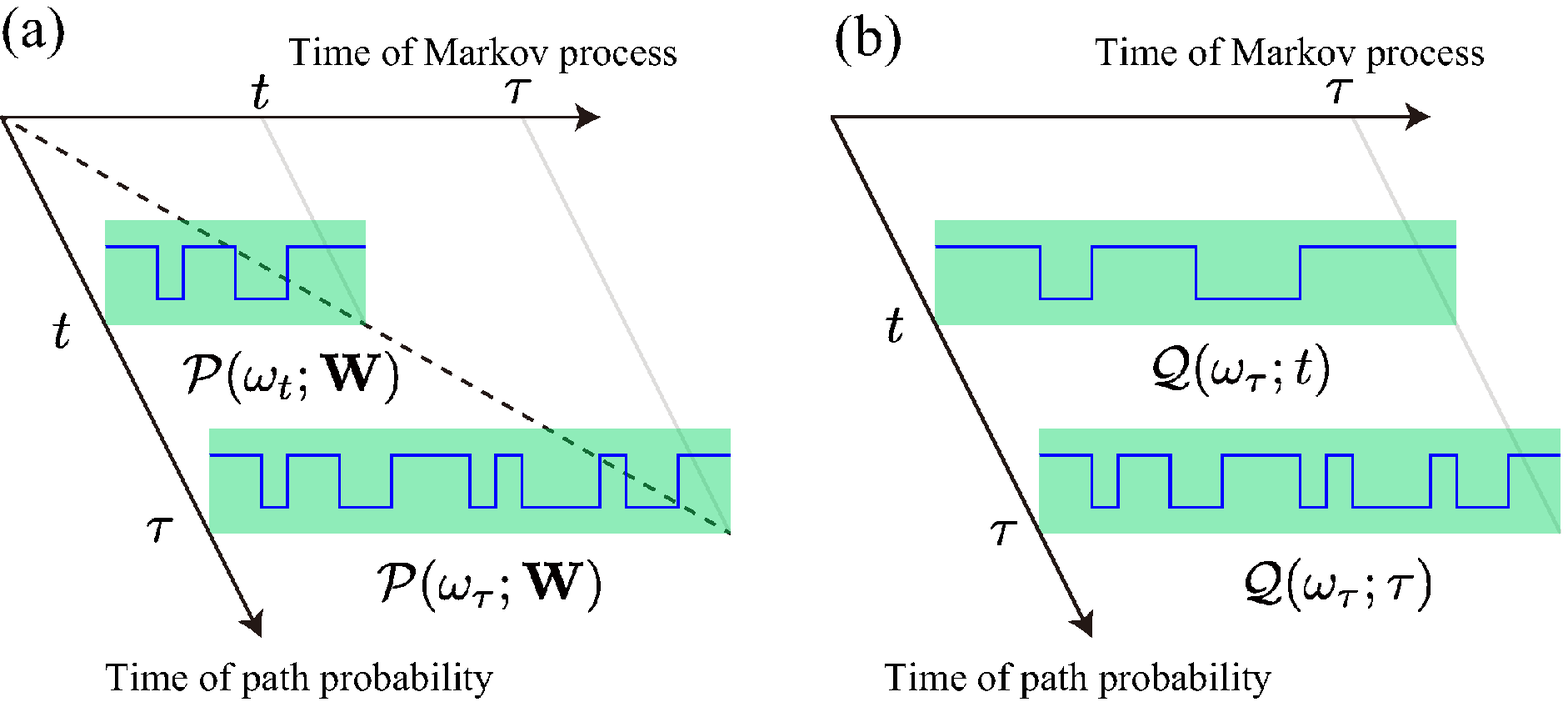} 
\caption{Intuitive illustration of the path probabilities. (a) Path probability $\mathcal{P}(\omega_t;\mathbf{W})$ and (b) scaled path probability $\mathcal{Q}(\omega_\tau;t)$. 
In (a) and (b), the horizontal direction denotes the time of the original Markov process while the downward direction denotes the time of the path probability. 
}
\label{fig:scaled_path}
\end{figure*}

Next, we consider a particular probability distribution. 
In speed limit relations, we often consider probability distribution $\mathbf{P}(t)$ or the time evolution of density operator $\rho(t)$. 
In contrast, we here consider the time evolution of the path probability. 
First, we fix the final time $\tau$ of the Markov process, where $\tau>0$.
Let $\omega_t = [X(t^\prime)]_{t^\prime = 0}^{t^\prime = t}$ be a stochastic trajectory of the Markov process within the interval $[0,t]$ where $0 \le t \le \tau$.
It is intuitive to consider the discretization of the interval $[0,t]$; 
therefore, we introduce a small lattice spacing $\Delta t$ for the discretization of $[0, t]$. 
Assuming that $t$ is divisible by $\Delta t$, the trajectory $\omega_t$ is 
\begin{align}
    \omega_{t}=[X(t^{\prime})]_{t^{\prime}=0}^{t^{\prime}=t}=[X(0),X(\Delta t),X(2\Delta t),\cdots,X(t-\Delta t),X(t)].
    \label{ea:omega_tau_def}
\end{align}
Thus, with discretization, the 
size
of $\omega_t$ are $t / \Delta t + 1$. 
Let $\mathcal{P}(\omega_\tau;\mathbf{W})$ be the probability of $\omega_\tau$ under transition rate $\mathbf{W}$.
As defined above, 
the Markov process begins at $0$ and ends at $\tau$. 
In the intermediate states of the time evolution, the Markov process
is defined for $[0,t]$ where $0 \le t \le \tau$. 
A set of intermediate states defined for $[0,t]$ exhibits a one-to-one correspondence with the path probability $\mathcal{P}(\omega_{t};\mathbf{W})$ [Fig.~\ref{fig:scaled_path}(a)]. 
Therefore, instead of considering the time evolution of the Markov process, we consider the time evolution of the path probability $\mathcal{P}(\omega_{t};\mathbf{W})$. 
However, 
as can be seen from Eq.~\eqref{ea:omega_tau_def},
the 
size
of $\omega_t$ depends on $t$, which prevents evaluation of the distance between $\mathcal{P}(\omega_t;\mathbf{W})$ at different $t$.
In particular, the Bhattacharyya coefficient $\mathrm{Bhat}\left(\mathcal{P}(\omega_{t_{1}};\mathbf{W}),\mathcal{P}(\omega_{t_{2}};\mathbf{W})\right)$ is not well defined for $t_1 \ne t_2$; hence, we can not calculate the similarity between path probabilities at different times. 
Instead, we introduce the following scaled path probability \cite{Hasegawa:2023:BulkBoundaryBoundNC}:
\begin{align}
    \mathcal{Q}(\omega_{\tau};t)\equiv\mathcal{P}\left(\omega_{\tau};\frac{t}{\tau}\mathbf{W}\right).
    \label{eq:mathcalQ_def}
\end{align}
We show the difference between $\mathcal{P}(\omega_t;\mathbf{W})$ and $\mathcal{Q}(\omega_t;t)$ 
in Figs.~\ref{fig:scaled_path}(a) and (b).
$\mathcal{Q}(\omega_\tau;t)$ is concerned with a Markov process whose dynamics are the same as those of the original process, except for its time scale.
The dynamics of $\mathcal{Q}(\omega_\tau;t)$ are $t/\tau$ times as fast as those in the original process. 
Therefore, the information at time $t$ in the original process with transition rate $\mathbf{W}$ can be obtained at time $\tau$ in the scaled process with transition rate $(t/\tau)\mathbf{W}$.
In $\mathcal{P}(\omega_t;\mathbf{W})$, the 
size
of $\omega_t$ is different for different $t$, which prevents calculation of the distance between $\mathcal{P}(\omega_t;\mathbf{W})$ at different $t$ values.
In contrast, because $\omega_\tau$ in $\mathcal{Q}(\omega_\tau;t)$ does not change, we can calculate the distance at different $t$. 

We now derive the upper bound for the total variation distance. 
From Eq.~\eqref{eq:TVD_Hel1},
the total variation distance is bounded from above by
\begin{align}
    \mathrm{TVD}(\mathcal{Q}(\omega_{\tau};t_{1}),\mathcal{Q}(\omega_{\tau};t_{2}))\leq\sqrt{1-\mathrm{Bhat}(\mathcal{Q}(\omega_{\tau};t_{1}),\mathcal{Q}(\omega_{\tau};t_{2}))^{2}}.
    \label{eq:TVD_Bhat_ineq}
\end{align}
By introducing an arbitrary observable $\mathcal{C}(\omega_\tau)$
that is a function of $\omega_\tau$, 
from Eqs.~\eqref{eq:F_Holder_ineq} and \eqref{eq:TVD_Bhat_ineq}, we obtain
\begin{align}
    \left|\braket{\mathcal{C}}_{t_{1}}-\braket{\mathcal{C}}_{t_{2}}\right|\le2\mathcal{C}_{\max}\sqrt{1-\mathrm{Bhat}(\mathcal{Q}(\omega_{\tau};t_{1}),\mathcal{Q}(\omega_{\tau};t_{2}))^{2}},
    \label{eq:obs_C_first_bound}
\end{align}
where $\mathcal{C}_{\max} \equiv \max_{\omega_\tau} |\mathcal{C}(\omega_\tau)|$ and the expectation $\braket{\mathcal{C}}_t$ should be evaluated with respect to $\mathcal{Q}(\omega_\tau;t)$. 
The right-hand side of Eq.~\eqref{eq:obs_C_first_bound} can be further bounded from above by the dynamical activity using the results of Ref.~\cite{Hasegawa:2023:BulkBoundaryBoundNC}.
As $\arccos\left[{\mathrm{Bhat}(\cdot,\cdot)}\right]$ constitutes the geodesic distance under the Fisher information metric \cite{Wootters:1981:StatDist},
the following relation holds \cite{Hasegawa:2023:BulkBoundaryBoundNC}:
\begin{equation}
    \frac{1}{2}\int_{t_{1}}^{t_{2}}\frac{\sqrt{\mathcal{A}(t)}}{t}dt\geq\arccos\left[\mathrm{Bhat}\left(\mathcal{Q}(\omega_{\tau};t_{1}),\mathcal{Q}(\omega_{\tau};t_{2})\right)\right],
    \label{eq:classical_bound}
\end{equation}
which yields
\begin{align}
    \cos\left[\frac{1}{2}\int_{t_{1}}^{t_{2}}\frac{\sqrt{\mathcal{A}(t)}}{t}dt\right]\le\mathrm{Bhat}\left(\mathcal{Q}(\omega_{\tau};t_{1}),\mathcal{Q}(\omega_{\tau};t_{2})\right),
    \label{eq:classical_bound2}
\end{align}
for $0 \le \frac{1}{2}\int_{t_{1}}^{t_{2}}\frac{\sqrt{\mathcal{A}(t)}}{t}dt \le \frac{\pi}{2}$. 
Substituting Eq.~\eqref{eq:classical_bound2} into Eq.~\eqref{eq:TVD_Bhat_ineq}, we obtain
\begin{align}
    \mathrm{TVD}(\mathcal{Q}(\omega_{\tau};t_{1}),\mathcal{Q}(\omega_{\tau};t_{2}))&\le\sqrt{1-\mathrm{Bhat}\left(\mathcal{Q}(\omega_{\tau};t_{1}),\mathcal{Q}(\omega_{\tau};t_{2})\right)^{2}}\nonumber\\&\le\sin\left[\frac{1}{2}\int_{t_{1}}^{t_{2}}\frac{\sqrt{\mathcal{A}(t)}}{t}dt\right].
    \label{eq:path_TVD_sin_ineq}
\end{align}
From Eqs.~\eqref{eq:obs_C_first_bound} and \eqref{eq:path_TVD_sin_ineq}, we obtain
\begin{align}
    \left|\braket{\mathcal{C}}_{t_{1}}-\braket{\mathcal{C}}_{t_{2}}\right|\le2\mathcal{C}_{\max}\sin\left[\frac{1}{2}\int_{t_{1}}^{t_{2}}\frac{\sqrt{\mathcal{A}(t)}}{t}dt\right],
    \label{eq:cMPS_operator_bound}
\end{align}
which holds true for $0\le\frac{1}{2}\int_{t_{1}}^{t_{2}}\frac{\sqrt{\mathcal{A}(t)}}{t}dt\le\frac{\pi}{2}$. 
Equation~\eqref{eq:cMPS_operator_bound} is the central inequality for deriving the thermodynamic correlation inequality. 

Now, we consider the calculation of $C(t)$. 
Using the path probability, the following relation is satisfied:
\begin{align}
    C(t)&=\braket{S(X(0))T(X(t))}\nonumber\\&=\sum_{\omega_{t}}\mathcal{P}(\omega_{t};\mathbf{W})S(X(0))T(X(t))\nonumber\\&=\sum_{\omega_{\tau}}\mathcal{P}\left(\omega_{\tau};\frac{t}{\tau}\mathbf{W}\right)S(X(0))T(X(\tau))\nonumber\\&=\sum_{\omega_{\tau}}\mathcal{Q}(\omega_{\tau};t)S(X(0))T(X(\tau)).
    \label{eq:Ct_calculation}
\end{align}
Equation~\eqref{eq:Ct_calculation} shows that the correlation function $C(t)$ can be computed using the scaled path probability $\mathcal{Q}(\omega_\tau;t)$.
Therefore, for the observable $\mathcal{C}(\omega_\tau)$, we employ the following function:
\begin{align}
    \mathcal{C}(\omega_{\tau})=S(X(0))T(X(\tau)),
    \label{eq:F_obs_def}
\end{align}
Then, from Eq.~\eqref{eq:Ct_calculation}, the correlation function $C(t)$ is the expectation of $\mathcal{C}(\omega_\tau)$ with respect to $\mathcal{Q}(\omega_\tau;t)$. That is,
$\braket{\mathcal{C}}_t = C(t)$. 
$\mathcal{C}_{\max}$ in Eq.~\eqref{eq:cMPS_operator_bound} is evaluated as
\begin{align}
    \mathcal{C}_{\max}=\max_{\omega_{\tau}}\left|\mathcal{C}(\omega_{\tau})\right|=\max_{\omega_{\tau}}\left|S(X(0))T(X(\tau))\right|=S_{\max}T_{\max}.
    \label{eq:F_Smax}
\end{align}
Combining Eq.~\eqref{eq:path_TVD_sin_ineq} and \eqref{eq:F_Smax} with Eq.~\eqref{eq:obs_C_first_bound}, 
we obtain Eq.~\mainUresultUAt{} in the main text. 

Let us derive the bound of Eq.~\mainUresultUeta{}. 
Instead of using Eq.~\eqref{eq:path_TVD_sin_ineq} to derive the bound, 
we consider the intermediate bound of Eq.~\eqref{eq:obs_C_first_bound} with $t_1 = 0$ and $t_2 =\tau$. 
The Bhattacharyya coefficient on the right-hand side of Eq.~\eqref{eq:obs_C_first_bound} yields
\begin{align}
    \mathrm{Bhat}\left(\mathcal{Q}\left(\omega_{\tau};0\right),\mathcal{Q}\left(\omega_{\tau};\tau\right)\right)=\sum_{\mu}P(\mu;0)\sqrt{e^{-\tau\sum_{\nu(\ne\mu)}W_{\nu\mu}}},
    \label{eq:Bhat_Q_calc}
\end{align}
the derivation of which is presented in Section~\ref{sec:Bhat_Q_calculation}.
Substituting Eq.~\eqref{eq:Bhat_Q_calc} into Eq.~\eqref{eq:obs_C_first_bound} gives Eq.~\mainUresultUeta{} in the main text. 

In Section~\ref{sec:tighter_bound}, we provide an improvement to $\mathcal{C}_{\mathrm{max}}$, which makes Eqs.~\mainUresultUAt{}, \mainUresultUzeroUtau{}, and \mainUresultUderivative{} tighter.

\section{Calculation of the Bhattacharyya coefficient\label{sec:Bhat_Q_calculation}}

In this section, we calculate $\mathrm{Bhat}\left(\mathcal{Q}\left(\omega_{\tau};0\right),\mathcal{Q}\left(\omega_{\tau};\tau\right)\right)$ in Eq.~\eqref{eq:Bhat_Q_calc}. 
We assume that a trajectory has $K$ jump events. 
Let $t_l$ be the timestamp of the $l$th jump,
where $l$ ranges from $0$ to $K$, $t_0$ is the start time ($t_0=0$), and $t_{K+1}$ is the end time ($t_{K+1}=\tau$). 
We define $B_{\mu_{l}}$ as the state following the $l$th jump. We begin with $B_{\mu_{0}}$, which represents the initial state of the process. The transition rate from $B_{\nu}$ to $B_{\mu}$ is denoted by $W_{\mu\nu}$. The initial distribution of the Markov process is denoted by $P(\mu_{0};0)$.
The path probability is given by
\begin{align}
    \mathcal{P}(\omega_{\tau};\mathbf{W})=P(\mu_{0};0)\mathcal{P}(\omega_{\tau}|\mu_{0};\mathbf{W}),
    \label{eq:full_path_prob}
\end{align}
where $\mathcal{P}(\omega_{\tau}|\mu_{0};\mathbf{W})$ is the conditional probability of $\omega_\tau$ given the initial state $B_{\mu_0}$:
\begin{align}
    \mathcal{P}(\omega_{\tau}|\mu_{0};\mathbf{W})=\prod_{l=1}^{K}W_{\mu_{l}\mu_{l-1}}e^{-\sum_{l=0}^{K}(t_{l+1}-t_{l})R(\mu_{l})},
    \label{eq:path_prob_def}
\end{align}
with
\begin{align}
    R(\mu)\equiv\sum_{\nu(\neq\mu)}W_{\nu\mu}.
    \label{eq:R_def}
\end{align}
The Bhattacharyya coefficient is expressed as
\begin{align}
    \mathrm{Bhat}(\mathcal{Q}(\omega_{\tau};0),\mathcal{Q}(\omega_{\tau};\tau))=\sum_{\omega_{\tau}}\sqrt{\mathcal{Q}(\omega_{\tau};0)\mathcal{Q}(\omega_{\tau};\tau)},
    \label{eq:Bhat_for_Q}
\end{align}
where $\mathcal{Q}(\omega_\tau;\tau)$ and $\mathcal{Q}(\omega_\tau;0)$, defined in Eq.~\eqref{eq:mathcalQ_def}, are given by
\begin{align}
    \mathcal{Q}(\omega_{\tau};\tau)&=P(\mu_{0};0)\prod_{l=1}^{K}W_{\mu_{l}\mu_{l-1}}e^{-\sum_{l=0}^{K}(t_{l+1}-t_{l})R(\mu_{l})},
    \label{eq:path_prob_Q_def}\\
    \mathcal{Q}(\omega_{\tau};0)&=P(\mu_{0};0)\times\begin{cases}
1 & K=0\\
0 & K\ge1
\end{cases}.
\label{eq:path_prob_Q_null_def}
\end{align}
For trajectories $\omega_\tau$ with $K=0$, $\mathcal{Q}(\omega_\tau;\tau)$ becomes
\begin{align}
    \mathcal{Q}(\omega_{\tau};\tau)=P(\mu_{0};0)e^{-\tau R(\mu_{0})}=P(\mu_{0};0)e^{-\tau\sum_{\nu(\ne\mu_{0})}W_{\nu\mu_{0}}}\,\,\,(K=0).
    \label{eq:Q_for_K0}
\end{align}
From Eq.~\eqref{eq:path_prob_Q_null_def}, 
we see that $\mathcal{Q}(\omega_\tau;0)$ is non-vanishing for $\omega_\tau$ with $K=0$, that is, trajectories with no jumps. 
Therefore, using Eq.~\eqref{eq:Q_for_K0}, the summation in Eq.~\eqref{eq:Bhat_for_Q} is evaluated as
\begin{align}
    \sum _{\omega _{\tau }}\sqrt{\mathcal{Q}( \omega _{\tau } ;0)\mathcal{Q}( \omega _{\tau } ;\tau )} &=\sum _{\omega _{\tau } ,K=0}\sqrt{\mathcal{Q}( \omega _{\tau } ;0)\mathcal{Q}( \omega _{\tau } ;\tau )}\nonumber\\
&=\sum _{\omega _{\tau } ,K=0} P(\mu ;0)\sqrt{e^{-\tau \sum _{\nu (\neq \mu _{0} )} W_{\nu \mu _{0}}}}\nonumber\\
&=\sum _{\mu } P(\mu ;0)\sqrt{e^{-\tau \sum _{\nu (\neq \mu _{0} )} W_{\nu \mu _{0}}}},
    \label{eq:Bhat_summation}
\end{align}
where
the last equality arises from understanding that the sum of the no-jump trajectories is equivalent to that of the initial probability distributions.
Equation~\eqref{eq:Bhat_summation} completes Eq.~\eqref{eq:Bhat_Q_calc}.

\section{Saturation of thermodynamic correlation inequality}

The main result [Eq.~\mainUresultUAt{}] involves three inequalities, Eqs.~\eqref{eq:F_Holder_ineq}, \eqref{eq:TVD_Bhat_ineq}, and \eqref{eq:path_TVD_sin_ineq}. 
Let us consider the upper bound for $|C(0)-C(t)|$, which
can be broken down into the following series of inequalities:
\begin{align}
    |C(0)-C(\tau)|&\le2S_{\mathrm{max}}T_{\mathrm{max}}\mathrm{TVD}(\mathcal{Q}(\omega_{\tau},0),\mathcal{Q}(\omega_{\tau},\tau))\label{eq:first_line}\\&\le2S_{\mathrm{max}}T_{\mathrm{max}}\sqrt{1-\mathrm{Bhat}(\mathcal{Q}(\omega_{\tau},0),\mathcal{Q}(\omega_{\tau},\tau))^{2}}\label{eq:second_line}\\&\le2S_{\mathrm{max}}T_{\mathrm{max}}\sin\left[\frac{1}{2}\int_{0}^{\tau}\frac{\sqrt{\mathcal{A}(t)}}{t}dt\right],
    \label{eq:third_line}
\end{align}
Equation~\mainUresultUzeroUtau{} is saturated when the equality conditions of the above three inequalities are satisfied simultaneously. 

The first line [Eq.~\eqref{eq:first_line}] is the H\"older inequality. 
Let us consider the equality condition for Eq.~\eqref{eq:F_Holder_ineq}.
Let $\mathcal{X}$ be a set defined as
\begin{align}
    \mathcal{X} \equiv \{x | F_{\mathrm{max}} = |F(x)|\}.
    \label{eq:X_def}
\end{align}
Consequently, the equality in Eq.~\eqref{eq:F_Holder_ineq} holds if and only if the following condition holds:
\begin{align}
    \left|p(x;t_{2})-p(x;t_{1})\right|=\begin{cases}
c_p & x\in\mathcal{X}\\
0 & x\notin\mathcal{X}
\end{cases},
\label{eq:Holder_eq_cond}
\end{align}
where $c_p$ denotes a positive real value. 
This condition implies that the probability difference is non-vanishing only at values of $x$ which yield the maximum $|F(x)|$ and that the difference is the same for all such $x$. 
 
We also consider the same equality condition for Eq.~\eqref{eq:first_line}.
We define
\begin{align}
    \Omega_\tau \equiv \{\omega_\tau | S_\mathrm{max}T_\mathrm{max} = |\mathcal{C}(\omega_\tau)|\}.
    \label{eq:Omega_def}
\end{align}
Consequently, the equality in Eq.~\eqref{eq:first_line} holds if and only if the following relation holds:
\begin{align}
    |\mathcal{Q}(\omega_{\tau};0)-\mathcal{Q}(\omega_{\tau};\tau)|=\begin{cases}
c_{\mathcal{Q}} & \omega_{\tau}\in\Omega_{\tau}\\
0 & \omega_{\tau}\notin\Omega_{\tau}
\end{cases},
\label{eq:Holder_eq_cond_omega}
\end{align}
where $c_\mathcal{Q}$ denotes a positive real value. 

Next, we consider the equality condition of the second line [Eq.~\eqref{eq:second_line}].
Let $\mathfrak{p}(x)$ and $\mathfrak{q}(x)$ be probability distributions.
The inequality of interest is given by Eq.~\eqref{eq:TVD_Hel1},
which is a direct consequence of the Cauchy-Schwartz inequality:
\begin{align}
    \mathrm{TVD}(\mathfrak{p},\mathfrak{q})^{2}&=\frac{1}{4}\left(\sum_{x}\left|\mathfrak{p}(x)-\mathfrak{q}(x)\right|\right)^{2}\nonumber\\&=\frac{1}{4}\left(\sum_{x}\left|\sqrt{\mathfrak{p}(x)}-\sqrt{\mathfrak{q}(x)}\right|\left(\sqrt{\mathfrak{p}(x)}+\sqrt{\mathfrak{q}(x)}\right)\right)^{2}\nonumber\\&\le\frac{1}{4}\left(\sum_{x}\left(\sqrt{\mathfrak{p}(x)}-\sqrt{\mathfrak{q}(x)}\right)^{2}\right)\left(\sum_{x}\left(\sqrt{\mathfrak{p}(x)}+\sqrt{\mathfrak{q}(x)}\right)^{2}\right)\nonumber\\&=1-\mathrm{Bhat}(\mathfrak{p},\mathfrak{q})^{2}.
    \label{eq:TVD_Bhat_ineq_detail}
\end{align}
As the equality condition for Eq.~\eqref{eq:TVD_Bhat_ineq_detail} directly follows that of the Cauchy-Schwartz inequality,
Eq.~\eqref{eq:TVD_Bhat_ineq_detail} achieves equality if and only if the following relation holds:
\begin{align}
    \left|\sqrt{\mathfrak{p}(x)}-\sqrt{\mathfrak{q}(x)}\right|=\mathfrak{c}\left(\sqrt{\mathfrak{p}(x)}+\sqrt{\mathfrak{q}(x)}\right)\hspace*{1em}(\mathrm{for\,all\,}x),
    \label{eq:equality_condition_TVD_Bhat}
\end{align}
where $\mathfrak{c}$ denotes a real proportionality coefficient. 
This corresponds to the following condition for Eq.~\eqref{eq:second_line}:
\begin{align}
    \left|\sqrt{\mathcal{Q}(\omega_{\tau};0)}-\sqrt{\mathcal{Q}(\omega_{\tau};\tau)}\right|=\mathfrak{c}\left(\sqrt{\mathcal{Q}(\omega_{\tau};0)}+\sqrt{\mathcal{Q}(\omega_{\tau};\tau)}\right)\hspace*{1em}(\mathrm{for\,all\,}\omega_{\tau}).
    \label{eq:equality_condition_TVD_Bhat_Q}
\end{align}
As there is no jump event at $t=0$, the following relation holds: 
\begin{align}
    \left|-\sqrt{\mathcal{Q}(\omega_{\tau};\tau)}\right|=\mathfrak{c}\sqrt{\mathcal{Q}(\omega_{\tau};\tau)}\hspace*{1em}(\mathrm{for}\,\omega_{\tau}\,\mathrm{with}\,K\ge1),
    \label{eq:equality_condition_TVD_Bhat_K1}
\end{align}
where $K$ denotes the number of jump events. 
Equation~\eqref{eq:equality_condition_TVD_Bhat_K1} immediately implies $\mathfrak{c}=1$, which, in turn, leads to
\begin{align}
    \mathcal{Q}(\omega_{\tau};\tau)=0\hspace*{1em}(\mathrm{for}\,\omega_{\tau}\,\mathrm{with}\,K=0).
    \label{eq:equality_condition_TVD_Bhat_K0}
\end{align}
The condition indicated by Eq.~\eqref{eq:equality_condition_TVD_Bhat_K0} is asymptotically satisfied when the value of $\tau$ approaches infinity. 
This is because, for $\tau \to \infty$, the probability of a no-jump trajectory approaches zero. 
Subsequently, satisfying the equality condition for Eq.~\eqref{eq:second_line} becomes increasingly plausible.

The third inequality [Eq.~\eqref{eq:third_line}] describes the relation between the path length connecting two points and their geodesic distance. 
In contrast to the equality conditions for Eqs.~\eqref{eq:first_line} and \eqref{eq:second_line},
identifying the equality condition for Eq.~\eqref{eq:third_line} is difficult. 
Suppose $\mathfrak{p}(x)$ and $\mathfrak{q}(x)$ are defined at $x = 1,2,\ldots,N$. 
We define two points:
\begin{align}
    \mathbf{v}_{P}\equiv\left[\begin{array}{c}
\sqrt{\mathfrak{p}(1)}\\
\sqrt{\mathfrak{p}(2)}\\
\vdots\\
\sqrt{\mathfrak{p}(N)}
\end{array}\right],\mathbf{v}_{Q}\equiv\left[\begin{array}{c}
\sqrt{\mathfrak{q}(1)}\\
\sqrt{\mathfrak{q}(2)}\\
\vdots\\
\sqrt{\mathfrak{q}(N)}
\end{array}\right].
\label{eq:vP_vQ_def}
\end{align}
The normalization condition of $\mathfrak{p}(x)$  is given by $\mathbf{v}_P\cdot \mathbf{v}_P=1$ (and similarly for $\mathbf{v}_Q$),
which shows that $\mathbf{v}_P$ and $\mathbf{v}_Q$ are located on an $N$-dimensional sphere. 
Therefore, the geodesic distance between $\mathfrak{p}(x)$ and $\mathfrak{q}(x)$ is $\arccos[{\mathbf{v}_P \cdot \mathbf{v}_Q}]$, which is the right-hand side of Eq.~\eqref{eq:classical_bound}. 
Let $\theta$ be a parameter satisfying $0\le \theta \le 1$. 
A point $\mathbf{v}(\theta)$, located on the geodesic between $\mathbf{v}_P$ and $\mathbf{v}_Q$, is expressed as
\begin{align}
    \mathbf{v}(\theta)\propto\theta\mathbf{v}_{P}+(1-\theta)\mathbf{v}_{Q}\hspace*{1em}(0\le\theta\le1),
    \label{eq:vR_def}
\end{align}
where the proportionality coefficient should be determined such that the norm of $\mathbf{v}(\theta)$ becomes unity, $\mathbf{v}(\theta)\cdot \mathbf{v}(\theta) = 1$. 
We cannot expect $\mathcal{Q}(\omega_\tau; t)$ to lie on the geodesic given by Eq.~\eqref{eq:vR_def}.
Hence, it may be challenging to determine whether the scaled path probability is closely aligned with the geodesic.

Consider the aforementioned equality conditions in the numerical simulations shown in Fig.~\FIGboundUplot{}(a).
Recall that the transition matrix and initial probability are
\begin{align}
    \mathbf{W}=\left[\begin{array}{cc}
0 & 1\\
0 & -1
\end{array}\right],\,\mathbf{P}(0)=\left[\begin{array}{c}
0\\
1
\end{array}\right].
\label{eq:W_P0_simulation}
\end{align}
Moreover, the score functions are 
\begin{align}
    \ensuremath{S(B_{1})=-1},\,\ensuremath{S(B_{2})=1},\,\ensuremath{T(B_{1})=-1},\,\ensuremath{T(B_{2})=1}.
    \label{eq:score_function_simulation}
\end{align}
$\Omega_\tau$ in Eq.~\eqref{eq:Omega_def} includes two varieties of trajectories: those that begin at $B_2$ and end at $B_2$, and those that begin at $B_2$ and end at $B_1$. When $\omega_\tau$ has $K=0$ (no jumps),  $|\mathcal{Q}(\omega_{\tau};0)-\mathcal{Q}(\omega_{\tau};\tau)|=|1-0|=1$
for $\tau \to \infty$. 
Moreover,
for $\tau \to \infty$,
the probability $P(1;\tau)$ approaches $1$, indicating that $|\mathcal{Q}(\omega_{\tau};0)-\mathcal{Q}(\omega_{\tau};\tau)|=|0-1|=1$ for $\omega_\tau$ beginning at $B_2$ and ending at $B_1$. This verifies that the first equality condition is satisfied when $\tau$ approaches $\infty$. The second equality condition can be readily fulfilled as $\tau$ approaches $\infty$, as explained previously.
Therefore, in Fig.~\FIGboundUplot{}(a), the first and second equality conditions are satisfied for $\tau \to \infty$. Although we cannot identify whether the third equality condition is satisfied, the right-hand side of Eq.~\mainUresultUzeroUtau{} becomes close to saturation as $\tau \to \infty$.

\section{Tighter bounds\label{sec:tighter_bound}}

In the main text, we derive several thermodynamic correlation inequalities [Eqs.~\mainUresultUAt{}--\mainUresultUeta{}]. 
These bounds can be tightened, as considered in Ref.~\cite{Hamazaki:2023:VSL}.
Recall that
\begin{align}
    \left|\braket{\mathcal{F}}_{t_{1}}-\braket{\mathcal{F}}_{t_{2}}\right|&=\left|\mathrm{Tr}\left[\mathcal{F}(\rho(t_{1})-\rho(t_{2}))\right]\right|\nonumber\\&=\left|\mathrm{Tr}\left[\left(\mathcal{F}-c\right)(\rho(t_{1})-\rho(t_{2}))\right]\right|\nonumber\\&\le2\left\Vert \mathcal{F}-c\right\Vert _{\mathrm{op}}\mathrm{TD}\left(\rho(t_{1}),\rho(t_{2})\right),
    \label{eq:Operator_ineq2}
\end{align}
where $c$ denotes an arbitrary real value. 
Therefore, the optimal upper bound can be obtained by minimizing the right-hand side of Eq.~\eqref{eq:Operator_ineq2} with respect to $c$. 
Subsequently, $F_\mathrm{max}$ in Eq.~\eqref{eq:LHS_classical} is replaced with
\begin{align}
    F_{\mathrm{max}}=\frac{1}{2}\left(\max_{x}F(x)-\min_{x}F(x)\right).
    \label{eq:Fmax_improved}
\end{align}
Therefore, $\mathcal{C}_\mathrm{max}$ in Eq.~\eqref{eq:F_Smax} is replaced with
\begin{align}
    \mathcal{C}_{\max}&=\frac{1}{2}\left(\max_{\omega_{\tau}}\mathcal{C}(\omega_{\tau})-\min_{\omega_{\tau}}\mathcal{C}(\omega_{\tau})\right)\nonumber\\&=\frac{1}{2}\left(\max_{\omega_{\tau}}S(X(0))T(X(\tau))-\min_{\omega_{\tau}}S(X(0))T(X(\tau))\right)\nonumber\\&=\frac{1}{2}\left(\max_{B_{1},B_{2}\in\mathcal{B}}S(B_{1})T(B_{2})-\min_{B_{1},B_{2}\in\mathcal{B}}S(B_{1})T(B_{2})\right).
    \label{eq:Cmax_tighten}
\end{align}
The main result in Eq.~\mainUresultUAt{} becomes
\begin{align}
    \left|C(t_{1})-C(t_{2})\right|\le\left(\max_{B_{1},B_{2}\in\mathcal{B}}S(B_{1})T(B_{2})-\min_{B_{1},B_{2}\in\mathcal{B}}S(B_{1})T(B_{2})\right)\sin\left[\frac{1}{2}\int_{t_{1}}^{t_{2}}\frac{\sqrt{\mathcal{A}(t)}}{t}dt\right].
\end{align}
This improvement can be applied to all inequalities derived from Eq.~\mainUresultUAt{}.

\section{Multi-point correlation function\label{sec:multipoint}}

Here, we show straight-forward generalizations of the thermodynamic correlation inequality presented in the main text. 

The two-point correlation function [Eq.~\correlationUdef{}] is the focus of the main text. 
It is easy to extend the bounds to the multipoint correlation functions. 
Consider a $J$-point correlation function ($J\ge 2$):
\begin{align}
\braket{S_{1}(t_{1})S_{2}(t_{2})\cdots S_{J}(t_{J})}\equiv\sum_{\nu_{1},\nu_{2},\cdots,\nu_{J}}S_{1}(B_{\nu_{1}})S_{2}(B_{\nu_{2}})\cdots S_{J}(B_{\nu_{J}})P(\nu_{1};t_{1})P(\nu_{2};t_{2}|\nu_{1};t_{1})\cdots P(\nu_{J};t_{J}|\nu_{J-1};t_{J-1}),
    \label{eq:multi_point_cor_def}
\end{align}
where $0 = t_1 < t_2 < \cdots < t_J$
and $S_i(B_\nu)$ ($i = 1,2,\ldots,J$) are score functions. 
The analogous relations of Eqs.~\mainUresultUzeroUtau{} and \mainUresultUeta{} for Eq.~\eqref{eq:multi_point_cor_def} can be obtained
by taking
\begin{align}
\mathcal{C}(\omega_{\tau})=S_{1}(X(t_{1}))S_{2}(X(t_{2}))\cdots S_{J}(X(t_{J})),
    \label{eq:C_omega_multipoint}
\end{align}
which provides the following bounds:
\begin{align}
    \left|\braket{S_{1}(0)S_{2}(0)\cdots S_{J}(0)}-\braket{S_{1}(t_{1})S_{2}(t_{2})\cdots S_{J}(t_{J})}\right|&\le2\left(\prod_{i=1}^{J}S_{i,\max}\right)\sqrt{1-\eta(t)}\label{eq:multipoint_bound2}\\&\le2\left(\prod_{i=1}^{J}S_{i,\max}\right)\sin\left[\frac{1}{2}\int_{0}^{t_{J}}\frac{\sqrt{\mathcal{A}(t^{\prime})}}{t^{\prime}}dt^{\prime}\right],\label{eq:multipoint_bound1}
\end{align}
where $S_{i,\mathrm{max}}\equiv\max_{B\in\mathcal{B}}|S_{i}(B)|$.

A bound for the one-point correlation function can also be obtained.
By considering $\mathcal{C}(\omega_\tau) = S(X(\tau))$, 
we obtain
\begin{align}
    |\braket{S(0)}-\braket{S(t)}|&\leq2S_{\max}\sqrt{1-\eta(t)}\label{eq:onepoint_bound2}\\&\leq2S_{\max}\sin\left[\frac{1}{2}\int_{0}^{t}\frac{\sqrt{\mathcal{A}\left(t^{\prime}\right)}}{t^{\prime}}dt^{\prime}\right],\label{eq:onepoint_bound1}
\end{align}
which correspond to a classical observable speed limit. 

Note that the one-point bound has other bounds using a classical speed limit
\cite{Vo:2022:UKTURJPA}. 
From Eq.~\eqref{eq:F_Holder_ineq}, we have
\begin{align}
    \left|\braket{S(0)}-\braket{S(t)}\right|&=\left|\sum_{\mu}S(X_{\mu})\left(P(\mu;0)-P(\mu;t)\right)\right|\\&\le2S_{\mathrm{max}}\mathrm{TVD}\left(\mathbf{P}(0),\mathbf{P}(t)\right).
    \label{eq:onepoint_alt_TVD}
\end{align}
Reference~\cite{Vo:2022:UKTURJPA} derived the following relation:
\begin{align}
    \mathrm{TVD}\left(\mathbf{P}(0),\mathbf{P}(t)\right)\le\mathcal{A}(t),
    \label{eq:S_alt_bound}
\end{align}
which yields
\begin{align}
    \left|\braket{S(0)}-\braket{S(t)}\right|\le2S_{\mathrm{max}}\mathcal{A}(t).
    \label{eq:S_alt_bound2}
\end{align}
Equation~\eqref{eq:S_alt_bound2} provides an alternative bound for $\left|\braket{S(0)}-\braket{S(t)}\right|$. 
The strength of Eq.~\eqref{eq:S_alt_bound2} lies in its ability to handle time-dependent transition rates, $W_{\nu\mu}(t)$. 
This contrasts with Eq.~\eqref{eq:onepoint_bound1}, which assumes a time-independent rate $W_{\nu\mu}$ (note that Eq.~\eqref{eq:onepoint_bound1} holds for non-steady-state dynamics). 
Equation~\eqref{eq:S_alt_bound2} is linear with respect to $t$ and is thus tighter than Eq.~\eqref{eq:onepoint_bound1} for a short time $t$,
while Eq.~\eqref{eq:onepoint_bound1} becomes tighter as $t \to \infty$ because the right-hand side of Eq.~\eqref{eq:onepoint_bound1} is bounded.

\section{Relation to thermodynamic uncertainty relation\label{sec:relation_to_TUR}}

We derived a thermodynamic uncertainty relation in Ref.~\cite{Hasegawa:2023:BulkBoundaryBoundNC}, which states
\begin{align}
    \left(\frac{\dblbrace{\mathcal{C}}_{\tau}+\dblbrace{\mathcal{C}}_{0}}{\braket{\mathcal{C}}_{\tau}-\braket{\mathcal{C}}_{0}}\right)^{2}\geq\frac{1}{\tan\left[\frac{1}{2}\int_{0}^{\tau}\frac{\sqrt{\mathcal{A}(t)}}{t}dt\right]^{2}},
    \label{eq:NC_TUR}
\end{align}
where $\mathcal{C}(\omega_\tau)$ is an observable of $\omega_\tau$ and $\dblbrace{\mathcal{C}}_{t}\equiv\sqrt{\braket{\mathcal{C}^{2}}_{t}-\braket{\mathcal{C}}_{t}^{2}}$.
Although the thermodynamic uncertainty relation derived in Ref.~\cite{Hasegawa:2023:BulkBoundaryBoundNC} concerns observables that count the number of jump events, Eq.~\eqref{eq:NC_TUR} can handle the observable that calculates the correlation $C(t)$,
as was defined in Eq.~\eqref{eq:F_obs_def}. 
Let us employ Eq.~\eqref{eq:F_obs_def} for $\mathcal{C}(\omega_\tau)$. 
The variance in Eq.~\eqref{eq:NC_TUR} can be evaluated as follows:
\begin{align}
   \dblbrace{\mathcal{C}}_{t}^{2}=\braket{\mathcal{C}^{2}}_{t}-\braket{\mathcal{C}}_{t}^{2}\le S_{\mathrm{max}}^{2}T_{\mathrm{max}}^{2}-\braket{\mathcal{C}}_{t}^{2}\le S_{\mathrm{max}}^{2}T_{\mathrm{max}}^{2}.
    \label{eq:F_var_upperbound}
\end{align}
Substituting Eq.~\eqref{eq:F_var_upperbound} into Eq.~\eqref{eq:NC_TUR}, we obtain
\begin{align}
    \left|C(0)-C(t)\right|\le2S_{\mathrm{max}}T_{\mathrm{max}}\tan\left[\frac{1}{2}\int_{0}^{t}\frac{\sqrt{\mathcal{A}(t^{\prime})}}{t^{\prime}}dt^{\prime}\right].
    \label{eq:Ct_bound_from_TUR}
\end{align}
Equation~\eqref{eq:Ct_bound_from_TUR} is similar to Eq.~\mainUresultUzeroUtau{}; however, the sine function in Eq.~\mainUresultUzeroUtau{} is replaced by a tangent function, indicating that Eq.~\mainUresultUzeroUtau{} is tighter than Eq.~\eqref{eq:Ct_bound_from_TUR}.

\section{Linear response\label{sec:Appendix_linear_response}}
Here, we present detailed calculations of the linear response theory.
Let us consider the application of a weak perturbation $\chi \mathbf{F} f(t)$ to the master equation \masterUequationUdef{}.
Considering the perturbation expansion with respect to $\chi$ up to the first order, the probability distribution is expanded as
\begin{align}
    \mathbf{P}(t)=\mathbf{P}_{st}+\chi\mathbf{P}_{1}(t),
    \label{eq:P_perturb_expansion}
\end{align}
where $\mathbf{P}_1(t)$ is a first-order term. 
Substituting Eq.~\eqref{eq:P_perturb_expansion} into Eq.~\masterUequationUdef{}, we obtain
\begin{align}
    \frac{d}{dt}\left(\mathbf{P}_{st}+\chi\mathbf{P}_{1}(t)\right)=\left(\mathbf{W}+\chi\mathbf{F}f(t)\right)\left(\mathbf{P}_{st}+\chi\mathbf{P}_{1}(t)\right),
    \label{eq:perturbed_master_eq}
\end{align}
in which collecting the terms with respect to  the order of $\chi$ yields
\begin{align}
    O(\chi^{0})\,\,\,\,&\frac{d}{dt}\mathbf{P}_{st}=\mathbf{W}\mathbf{P}_{st},\label{eq:P0_ODE}\\
    O(\chi^{1})\,\,\,\,&\frac{d}{dt}\mathbf{P}_{1}(t)=\mathbf{W}\mathbf{P}_{1}(t)+\mathbf{F}\mathbf{P}_{st}f(t).
    \label{eq:P1_ODE}
\end{align}
The zeroth order equation [Eq.~\eqref{eq:P0_ODE}] vanishes by definition because $\mathbf{P}_{st}$ is assumed to be the steady-state solution of Eq.~\masterUequationUdef{}.
$\mathbf{P}_1(t)$ is given by Eq.~\PIUsolUmain{} in the main text, 
in which we consider the score function $G(B_\nu)$ and its expectation. 
The change in $\braket{G}$ due to the perturbation can be represented by
\begin{align}
    \Delta G(t)&\equiv\chi\mathds{1}\mathbf{G}\mathbf{P}_{1}(t)\nonumber\\
    &=\chi\int_{-\infty}^{t}\mathds{1}\mathbf{G}e^{\mathbf{W}(t-t^{\prime})}\mathbf{F}\mathbf{P}_{st}f(t^{\prime})dt^{\prime}\nonumber\\
    &=\chi\int_{-\infty}^{\infty}R_{G}(t-t^{\prime})f(t^{\prime})dt^{\prime},
    \label{eq:lienar_response_R}
\end{align}
where $R_{G}(t)$ denotes the linear response function [Eq.~\linearUresponseUfunctionUdef{}]. 
From Eq.~\correlationUdef{}, the time derivative of $C(t)$ is
\begin{align}
    \frac{d}{dt}C(t)=\mathds{1}\mathbf{T}e^{\mathbf{W}t}\mathbf{W}\mathbf{S}\mathbf{P}_{st}.
    \label{eq:dC_dtau}
\end{align}
In the main text, we consider the case 
$\mathbf{G}=\mathbf{T}$
and $\mathbf{F}=\mathbf{W}\mathbf{S}$, which is assumed in the following. 
The perturbation $\mathbf{W}\mathbf{S}$ can be expressed as
\begin{align}
    \mathbf{W}\mathbf{S}=\left[\begin{array}{cccc}
S_{11}W_{11} & S_{22}W_{12} & \cdots & S_{NN}W_{1N}\\
S_{11}W_{21} & S_{22}W_{22} &  & S_{NN}W_{2N}\\
\vdots &  & \ddots & \vdots\\
S_{11}W_{N1} & S_{22}W_{N2} & \cdots & S_{NN}W_{NN}
\end{array}\right].
\label{eq:WS_expression}
\end{align}
We immediately obtain
\begin{align}
    R_{T}(t)=\frac{d}{dt}C(t).
    \label{eq:R_equal_Ctau}
\end{align}

For the pulse perturbation $f(t)=\delta(t)$, where $\delta(t)$ is the Dirac delta function, we obtain
\begin{align}
    \Delta T^{(p)}(t)&=\chi\int_{-\infty}^{\infty}R_{T}(t-t^{\prime})\delta(t^{\prime})dt^{\prime}\nonumber\\&=\chi R_{T}(t)\nonumber\\&=\chi\frac{d}{dt}C(t).
    \label{eq:Delta_S_as_Ctau}
\end{align}
Using Eq.~\mainUresultUderivative{}, we obtain Eq.~\DeltaUSUpUbound{}. 

Next, we consider the step perturbation, that is, $f(t)=\Theta(t)$, where $\Theta(t)$ is the Heaviside step function:
\begin{align}
    \Theta(t)=\begin{cases}
0 & (t<0)\\
1 & (t\ge0)
\end{cases}.
\label{eq:Heaviside_function_def}
\end{align}
Then we have
\begin{align}
    \Delta T^{(s)}(t)&=\chi\int_{-\infty}^{\infty}R_{T}(t-t^{\prime})\Theta(t^{\prime})dt^{\prime}\nonumber\\&=\chi\int_{0}^{t}R_{T}(t-t^{\prime})dt^{\prime}\nonumber\\&=\chi\int_{0}^{t}R_{T}(t^{\prime})dt^{\prime}\nonumber\\&=\chi\int_{0}^{t}\frac{dC(t^{\prime})}{dt^{\prime}}dt^{\prime}\nonumber\\&=\chi\left(C(t)-C(0)\right),
    \label{eq:response_step}
\end{align}
which yields Eq.~\DeltaUSUsUbound{} in the main text.

\section{Norm and distance measures}

Here, for the readers' convenience, we review the norm and distance measures for quantum and classical systems. 
Let $A$ and $B$ be arbitrary Hermitian operators. The 
Schatten
$p$-norm is defined as
\begin{align}
    \left\Vert A\right\Vert _{p}\equiv\left[\mathrm{Tr}\left[\left(\sqrt{A^2}\right)^{p}\right]\right]^{\frac{1}{p}}=\left(\sum_{\lambda\in\mathrm{evals}(A)}|\lambda|^{p}\right)^{\frac{1}{p}},
    \label{eq:Shattan_norm_def}
\end{align}
where $\mathrm{evals}(A)$ provides the set of eigenvalues for $A$. 
For a particular $p$, we have
\begin{align}
    \left\Vert A\right\Vert _{\mathrm{op}}&=\left\Vert A\right\Vert _{\infty}=\max_{\lambda\in\mathrm{evals}(A)}|\lambda|,\label{eq:Shatten_inf}\\
    \left\Vert A\right\Vert _{\mathrm{tr}}&=\left\Vert A\right\Vert _{1}=\mathrm{Tr}\left[\sqrt{A^{2}}\right],\label{eq:Shattan_one}\\
    \left\Vert A\right\Vert _{\mathrm{hs}}&=\left\Vert A\right\Vert _{2}=\sqrt{\mathrm{Tr}\left[A^{2}\right]}.\label{eq:Shattan_two}
\end{align}
Equations~\eqref{eq:Shatten_inf}, \eqref{eq:Shattan_one}, and \eqref{eq:Shattan_two} are referred to as the operator, trace, and Hilbert--Schmidt norms, respectively. 
The H\"older inequality states
\begin{align}
    \left|\mathrm{Tr}\left[AB\right]\right|\le\left\Vert A\right\Vert _{p}\left\Vert B\right\Vert _{q}.
    \label{eq:matrix_Holder_ineq}
\end{align}
where $p$ and $q$ must satisfy the condition $1/p + 1/q = 1$. 
When $p=q=2$, Eq.~\eqref{eq:matrix_Holder_ineq} reduces to the Cauchy-Schwarz inequality. 
In particular, we use the case when $p=\infty$ and $q=1$:
\begin{align}
    \left|\mathrm{Tr}\left[AB\right]\right|\le\left\Vert A\right\Vert _{\mathrm{op}}\left\Vert B\right\Vert _{\mathrm{tr}}.
    \label{eq:Holder_inf_1}
\end{align}

We define the trace distance and quantum fidelity as
\begin{align}
    \mathrm{TD}(\rho,\sigma)&\equiv\frac{1}{2}\left\Vert \rho-\sigma\right\Vert _{1},\label{eq:trace_dist_def}\\
    \mathrm{Fid}(\rho,\sigma)&\equiv\left(\mathrm{Tr}\sqrt{\sqrt{\rho}\sigma\sqrt{\rho}}\right)^{2}.
    \label{eq:fidelity_def}
\end{align}
When considering the pure states $\ket{\psi}$ and $\ket{\phi}$, the fidelity and trace distance are reduced to
\begin{align}
\mathrm{Fid}\left(\ket{\psi},\ket{\phi}\right)&=\left|\braket{\psi|\phi}\right|^{2},\label{eq:Fid_pure_state}\\
    \mathrm{TD}(\ket{\psi},\ket{\phi})&=\sqrt{1-|\braket{\psi|\phi}|^{2}}.
    \label{eq:TD_pure_state}
\end{align}
These two distances are related as follows:
\begin{align}
    \mathrm{TD}(\rho,\sigma)\le\sqrt{1-\mathrm{Fid}(\rho,\sigma)}.
    \label{eq:TD_Fid_ineq}
\end{align}
The equality of Eq.~\eqref{eq:TD_Fid_ineq} holds true when both $\rho$ and $\sigma$ are pure \cite{Nielsen:2011:QuantumInfoBook}. 

Here, we introduce the related classical probability distance measures. 
Let $\mathfrak{p}(x)$ and $\mathfrak{q}(x)$ be probability distributions. 
The total variation and Hellinger distances are defined as follows: 
\begin{align}
    \mathrm{TVD}(\mathfrak{p},\mathfrak{q})&\equiv \frac{1}{2}\sum_{x}|\mathfrak{p}(x)-\mathfrak{q}(x)|,\label{eq:TVD_def}\\
    \mathrm{Hel}^{2}(\mathfrak{p},\mathfrak{q})&\equiv\frac{1}{2}\sum_{x}\left(\sqrt{\mathfrak{p}(x)}-\sqrt{\mathfrak{q}(x)}\right)^{2}\label{eq:Hellinger_def}\\
    &=1-\mathrm{Bhat}(\mathfrak{p},\mathfrak{q})
    \label{eq:Hel_Bhat_relation}
\end{align}
where $\mathrm{Bhat}(\mathfrak{p},\mathfrak{q})$ denotes the Bhattacharyya coefficient:
\begin{align}
    \mathrm{Bhat}\left(\mathfrak{p},\mathfrak{q}\right)\equiv\sum_{x}\sqrt{\mathfrak{p}(x)\mathfrak{q}(x)}.
    \label{eq:Bhat_coef_def}
\end{align}
The following relations hold between the total variation and Hellinger distances \cite{LeCam:1973:Convergence,Sason:2016:DivIneqReview}:
\begin{align}
    \mathrm{Hel}^{2}(\mathfrak{p},\mathfrak{q})&\leq\mathrm{TVD}(\mathfrak{p},\mathfrak{q})\label{eq:Hel_TVD}\\
    &\leq\sqrt{\mathrm{Hel}^{2}(\mathfrak{p},\mathfrak{q})(2-\mathrm{Hel}^{2}(\mathfrak{p},\mathfrak{q}))}\label{eq:TVD_Hel1}\\
    &\leq\sqrt{2\mathrm{Hel}^{2}(\mathfrak{p},\mathfrak{q})}.
    \label{eq:TVD_Hel2}
\end{align}